\documentclass[prb,twocolumn,showpacs,preprintnumbers,amsmath,amssymb]{revtex4}
\usepackage{graphicx}
\usepackage{longtable}
\usepackage{dcolumn}
\usepackage{bm}
\usepackage{color}

\providecommand{\mr}{\mathrm}

\begin{document}

\title{\emph{Ab-initio} calculation of the effective on-site Coulomb interaction
parameters \\ for half-metallic magnets}

\author{Ersoy \c{S}a\c{s}{\i}o\u{g}lu$^{1}$}\email{e.sasioglu@fz-juelich.de}
\author{Iosif Galanakis$^{2}$}\email{galanakis@upatras.gr}
\author{Christoph Friedrich$^{1}$}
\author{Stefan  Bl\"{u}gel$^{1}$}

\affiliation{$^{1}$Peter Gr\"{u}nberg Institut and Institute for
Advanced Simulation, Forschungszentrum J\"{u}lich and JARA, D-52425
J\"{u}lich, Germany\\
$^{2}$Department of Materials Science, School of Natural Sciences,
University of Patras,  GR-26504 Patra, Greece}

\date{\today}

\begin{abstract}

Correlation effects play an important role in the electronic
structure of half-metallic (HM) magnets. In particular, they give
rise to non-quasiparticle states above (or below) the Fermi energy
at finite temperatures that reduce the spin polarization and, as a
consequence, the efficiency of spintronics devices. Employing the
constrained random-phase approximation (cRPA) within the
full-potential linearized augmented-plane-wave (FLAPW) method
using maximally localized Wannier functions, we calculate the
strength of the effective on-site Coulomb interaction (Hubbard $U$
and Hund exchange $J$) between localized  electrons in different
classes of HM magnets considering: (i) \emph{sp}-electron
ferromagnets in rock-salt structure, (ii) zincblende 3\emph{d}
binary ferromagnets, as well as (iii) ferromagnetic and
ferrimagnetic semi- and full-Heusler compounds. For HM
\emph{sp}-electron ferromagnets, the calculated Hubbard $U$
parameters are between 2.7 eV and 3.9 eV, while for
transition-metal-based HM compounds they lie between 1.7 and 3.8
eV, being smallest for MnAs (Mn-3\emph{d} orbitals) and largest
for Cr$_2$CoGa (Co-3\emph{d} orbitals). For the HM full-Heusler
compounds, the  Hubbard $U$ parameters are comparable to the ones
in elementary 3\emph{d} transition metals, while for semi-Heusler
compounds they are slightly smaller. We show that the increase of
the Hubbard $U$ with structural complexity, i.e., from MnAs to
Cr$_2$CoGa, stems from the screening of the \emph{p} electrons of
the non-magnetic \emph{sp} atoms. The \emph{p}-electron screening
turns out to be more efficient for MnAs than for Cr$_2$CoGa. The
calculated Hubbard $U$ parameters for CrAs, NiMnSb, and Co$_2$MnSi
are about two times smaller than previous estimates based on the
cLDA method. Furthermore, the width of the correlated \emph{d} or
\emph{p} bands of the studied compounds is usually smaller than
the calculated Hubbard $U$ parameters. Thus, these HM magnets
should be classified as weakly correlated materials.

\end{abstract}


\maketitle

\section{Introduction}
\label{sec1}

The field of spintronics is one of the most rapidly expanding
fields of nanoscience and technology because the incorporation of
the electron's spin offers an additional degree of freedom to be
used for information processing in nanodevices.\cite{ReviewSpin} A
key role in this research field is played by \textit{ab-initio}
studies of the electronic structure within density functional
theory (DFT), which have allowed the modelling of the properties
of several materials prior to their experimental growth.  Among
the materials that might find application in future magnetic
nanodevices are the half-metallic (HM)
magnets.\cite{ReviewHM,Review_DMS}

The ferromagnetic semi-Heusler compound NiMnSb was the first
material for which the HM character was predicted and
described;\cite{Groot1983} it exhibits usual metallic behavior for
one spin direction, while an energy gap in the band structure is
present in the other spin direction as in semiconductors. The
prospect of creating 100\% spin-polarized current has triggered
interest in such compounds and since the initial prediction of de
Groot \textit{et al.} several HM compounds have been
discovered.\cite{Pickett07,FelserRev} Several aspects concerning
the implementation of HM alloys in realistic devices, like
magnetic tunnel junctions, spin valves, and spin transistors, have
been discussed in
literature.\cite{Chadov11,Lezaic06,Mavropoulos05,Spin_Transistor}

For HM magnets, mean-field calculations of the electronic
structure, such as DFT, yield 100\% spin polarization at the Fermi
level. However, correlation effects among the localized electrons
lead to the appearance of non-quasiparticle states above (or
below) the Fermi level at finite temperatures. These states stem
from the electron-magnon interaction and cannot be described
within DFT irrespective of the correlation strength. The existence
of these states has been experimentally confirmed by recent
magnetic tunnel junction spectroscopy measurements on the
ferromagnetic HM full-Heusler compound
Co$_2$MnSi.\cite{ChioncelPRL08} Non-quasiparticle states severely
affect the perfect spin polarization above (or below) the Fermi
level degrading the performance of spintronics devices. Moreover
their behavior is material specific and thus extensive
calculations of the electronic structure of half-metals are needed
including correlation effects.

Electronic structure calculations based on DFT with local or
semilocal approximations for the exchange-correlation functional
are quite successful for materials from weak to intermediate
electronic correlations. However, they fail for systems with
strong electronic correlations. There are two common ways to
include correlations in first-principles electronic structure
calculations. The first one is the so-called LDA+$U$ scheme, in
which the local-density approximation (LDA) of DFT is augmented by
an on-site Coulomb repulsion term and an exchange term with the
Hubbard $U$ and Hund exchange $J$ parameters,
respectively.\cite{JepsenPRB2010,SolovyevReview} Such a scheme has
been applied for example to Co$_2$FeSi, showing that correlations
restore the HM character of the compound,\cite{FelserU} and to
NiMnSb.\cite{ChioncelPRB06_2} But LDA+$U$ cannot describe the
non-quasiparticle states. A more elaborate modern computational
scheme, which combines many-body model Hamiltonian methods with
DFT, is the so-called LDA+DMFT method, where DMFT stands for
Dynamical Mean-Field Theory.\cite{MinarReview,DMFT} In this
scheme, the interacting many-body system is mapped onto the
subspace of localized states, formed by $d$ or $p$ orbitals in the
present compounds, where the interaction with the rest of the
system is again incorporated in a Hubbard $U$ and Hund exchange
$J$ parameter. The still very complex many-body problem in the
correlated subspace is solved as an Anderson impurity problem
embedded in a dynamical mean field---in the form of a frequency
dependent self-energy---that accounts for all other sites.
LDA+DMFT has been applied to several HM magnetic systems like
Co$_2$MnSi,\cite{ChioncelPRL08}
NiMnSb,\cite{ChioncelPRB03,ChioncelPRB06,ChioncelPRB10}
FeMnSb,\cite{ChioncelPRL06} Mn$_2$VAl,\cite{ChioncelMn2VAl}
VAs\cite{ChioncelVAs} and CrAs.\cite{ChioncelCrAs,ChioncelCrAs2}
Indeed, in all these compounds the LDA+DMFT method yielded
non-quasiparticle states above (or below) the Fermi energy.

Thus, the Coulomb interaction parameters (Hubbard $U$ and Hund
exchange $J$) play a crucial role in the study of the correlation
effects in solids. However, their determination from experimental
data is a difficult task, which impedes the predictive power of
these approaches. Therefore, a direct calculation of these
parameters in solids from first principles is highly desirable.
Several authors have addressed this problem and a number of
different approaches have been proposed and applied to the bulk
phase of various classes of
materials.\cite{Kotani,Solovyev,Schnell,cLDA1,cLDA2,cLDA3,cRPA,cRPA_2,cRPA_3,Wehling,cRPA_Sasioglu,Zhang_Brothers,Hunter,Kaltak}
Among them, the constrained local-density approximation (cLDA) is
the most popular,\cite{cLDA1,cLDA2,cLDA3} but cLDA is known to
give unreasonably large Hubbard $U$ values for the late transition
metal atoms due to difficulties in compensating for the
self-screening error of the localized electrons.\cite{cRPA_2} On
the other hand, the constrained random-phase approximation (cRPA),
though numerically much more demanding, does not suffer from these
difficulties and offers an efficient way to calculate the
effective Coulomb interaction parameters in solids. Moreover, cRPA
allows to determine individual Coulomb matrix elements, e.g., on
site, off site, intra-orbital, inter-orbital, and exchange, as
well as their frequency dependence.\cite{cRPA,cRPA_Sasioglu}

Despite enormous work on HM magnets, no determination of the
Coulomb interaction parameters exists within the cRPA approach.
Available cLDA calculations of Hubbard $U$ parameter for HM CrAs,
NiMnSb, and Co$_2$MnSi turned out to be unreasonably large (6-7
eV). Thus, previous studies employing the Hubbard $U$ either
assume values close to the ones of the elementary transition metal
(TM) atoms or are performed for a variety of Hubbard $U$
values.\cite{FelserU} On the other hand, previous cRPA
calculations for TMs have shown that the Hubbard $U$ values are
sensitive to a variety of factors like the crystal structure, the
spin-polarization, the \emph{d} electron number and the \emph{d}
orbital filling,\cite{cRPA_Sasioglu} and thus values for the
elementary TMs cannot be directly used for complex intermetallic
compounds. The aim of the present work is to present a systematic
study of the effective on-site Coulomb interaction parameters
(Hubbard $U$ and Hund exchange $J$) between localized \emph{d} or
\emph{p} electrons in 20 HM magnets. We consider representatives
of the (i) semi-Heusler compounds like NiMnSb, (ii) ferrimagnetic
full-Heusler compounds like Mn$_2$VAl, (iii) inverse full-Heusler
compounds like Cr$_2$CoGa, (iv) usual L$2_1$-type ferromagnetic
full-Heusler compounds, (v) transition-metal pnictides like CrAs,
and finally (vi) \emph{sp}-electron (also called \emph{d}$^0$)
ferromagnets like CaN. Thus, our study covers a wide range of HM
magnets allowing for a deeper understanding of the behavior of the
Coulomb interactions parameters of the same element in different
HM magnetic systems. To calculate the effective Coulomb
interaction parameters, we have employed the cRPA method within
the full-potential linearized augmented-plane-wave (FLAPW) method
using maximally localized Wannier functions (MLWFs).

The paper is organized as follows. In Section\,\ref{sec2}, we
shortly present the methodology behind cRPA calculations. In
Section\,\ref{sec3}, we present calculated values
of Coulomb interaction parameters for a variety of well-known HM
magnets. Finally, we summarize our conclusions in Section\,\ref{sec4}.

\section{Computational Method}
\label{sec2}

\begin{figure}[t]
\begin{center}
\includegraphics[width=\columnwidth]{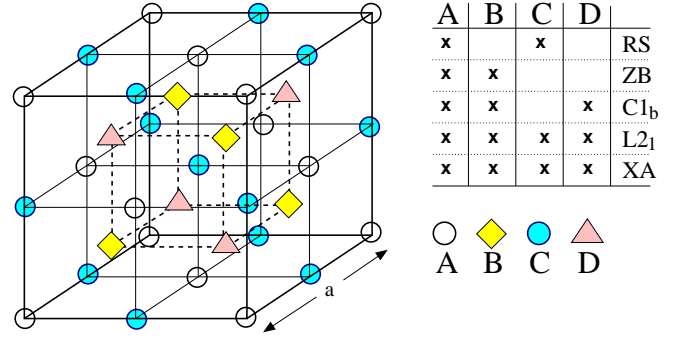}
\end{center}
\vspace*{-0.6 cm}\caption{(color online) Schematic representation
of the cubic structure of the various lattices adopted by the
present compounds. The cube contains exactly four primitive unit
cells.} \label{fig1}
\end{figure}

All compounds considered in this paper crystallize in a cubic
structure as shown in Fig.\,\ref{fig1}. The lattice consists of
four interpenetrating fcc lattices with Wyckoff positions:
$\mathrm{A}=(0\: 0\: 0)$, $\mathrm{B}=(\frac{1}{4}\: \frac{1}{4}\:
\frac{1}{4})$, $\mathrm{C}=(\frac{1}{2}\: \frac{1}{2}\:
\frac{1}{2})$, and $\mathrm{D}=(\frac{3}{4}\: \frac{3}{4}\:
\frac{3}{4})$. In the case of the rock-salt (RS) [zincblende (ZB)]
structure, the B and D (C and D) sites are vacant. In the
C$1_b$-type structure adapted by the semi-Heusler compounds (XYZ),
the X,Y, and Z atoms occupy the A, B, and D sites, respectively,
and the C site is vacant. Full-Heusler compounds possess either
the L$2_1$-type or the XA-type structure depending on the valency
of the X and Y elements. If the valency of the X elements is
larger (smaller) than that of the Y element, the compound prefers
the L$2_1$- (XA)-type structure. In the XA-type structure, the
unit cell is occupied in the sequence X-X-Y-Z instead of the
X-Y-X-Z sequence in the L$2_1$-type structure; the two X atoms are
not anymore equivalent. For the HM Heusler compounds as well as
for some zincblende systems, we have used experimental lattice
parameters, while for the \textit{sp}-electron ferromagnets
theoretical ones are used (see Table\,\ref{table1}). The
ground-state calculations are carried out using the FLAPW method
as implemented in the \texttt{FLEUR} code \cite{Fleur} with the
generalized gradient approximation (GGA) to the
exchange-correlation potential as parameterized by Perdew
\textit{et al.}\cite{GGA} A dense $16 \times 16 \times 16$
$\mathbf{k}$-point grid is used to perform the numerical
integrations in the Brillouin zone. The maximally localized
Wannier functions (MLWFs) are constructed with the
\texttt{Wannier90} code.\cite{Max_Wan,Wannier90,Fleur_Wannier90}
The effective Coulomb potential is calculated within the recently
developed cRPA method \cite{cRPA,cRPA_2,cRPA_3} implemented in the
\texttt{SPEX} code \cite{Spex} (for further technical details see
Refs.\,\onlinecite{cRPA_Sasioglu} and \onlinecite{Sasioglu}). We
use an $8 \times 8 \times 8$ $\mathbf{k}$-point grid in the cRPA
calculations. In the rest of the section, we will sketch the
formalism used to calculate the effective Coulomb potential.

In the present work, the correlated \emph{d} or \emph{p} subspace
is spanned by a set of $\textrm{MLWFs}$, which are given by
\begin{equation}
w_{n\textbf{R}}^{\sigma}(\textbf{r}) =
\frac{1}{N}\sum_{\mathbf{k}}e^{-i\mathbf{k}\cdot\textbf{R}}\sum_{m}T_{\mathbf{R},mn}^\sigma(\mathbf{k})
\varphi_{\mathbf{k}m}^{\sigma}(\textbf{r})\,, \label{Wannier}
\end{equation}
where $N$ is the number of $\mathbf{k}$ points,
$T_{\mathbf{R},mn}^{{\sigma}(\mathbf{k})}$ is the unitary
transformation matrix,
$\varphi_{{\mathbf{k}}m}^{\sigma}(\mathbf{r})$ are single-particle
Kohn-Sham states of spin $\sigma$ and band index $m$, and
$\mathbf{R}$ is the atomic position vector in the unit cell. The
transformation matrix $T_{mn}^{\sigma(\mathbf{k})}$ is determined
by minimizing the spread
\begin{equation}
\Omega=\sum_{n,\sigma}\big(\langle
w_{n\mathbf{0}}^{\sigma}|r^{2}|w_{n\mathbf{0}}^{\sigma} \rangle
-\langle
w_{n\mathbf{0}}^{\sigma}|\textbf{r}|w_{n\mathbf{0}}^{\alpha}
\rangle^2\big)\,, \label{spread}
\end{equation}
where the sum runs over all Wannier functions. We choose the
\textit{p} states as our correlated subspace for Rocksalt
\textit{sp}-electron materials, while for zincblende and Heusler
compounds the \textit{d} states form the correlated subspace.

Within the RPA, the polarization function is written as
\begin{eqnarray} \nonumber
P(\mathbf{r},\mathbf{r^{\prime}};\omega)&=&\sum_{\sigma}\sum_{\mathbf{k},m}^{\mr
{occ}} \sum_{\mathbf{k}^\prime,m^{\prime}}^{\mr {unocc}}
\varphi_{\mathbf{k}m}^{\sigma}(\mathbf{r}) \varphi_{\mathbf{k}^\prime m^{\prime}}^{\sigma
*}(\mathbf{r}) \varphi_{\mathbf{k}m}^{\sigma *}(\mathbf{r^{\prime}})
\varphi_{\mathbf{k}^\prime m^{\prime}}^{\sigma}(\mathbf{r^{\prime}}) \\  &&\times \bigg
[\frac{1}{\omega - \Delta_{\mathbf{k}m,\mathbf{k}^\prime m^{\prime}}^{\sigma}} -
\frac{1}{\omega + \Delta_{\mathbf{k}m,\mathbf{k}^\prime m^{\prime}}^{\sigma}}\bigg ]
\label{polar}
\end{eqnarray}
with $\Delta_{\mathbf{k}m,\mathbf{k}^\prime
m^{\prime}}^{\sigma}=\epsilon_{\mathbf{k}^\prime
m^{\prime}}^{\sigma} - \epsilon_{\mathbf{k}m}^{\sigma}-i\delta$,
the Kohn-Sham eigenvalues $\epsilon_{\mathbf{k}m}^\sigma$, and a
positive infinitesimal $\delta$. The $\sigma$ runs over both spin
channels. The basic idea of the cRPA is to define an effective
interaction $U$ between the localized (correlated) electrons by
restricting the screening processes to those that are not
explicitly treated in the effective model Hamiltonian. To this
end, we divide the full polarization matrix $P = P_{\mr l} +
P_{\mr r}$, where $P_{\mr l}$ includes transitions only between
localized states and $P_{\mr r}$ is the remainder.

For the calculation of the polarization matrix $P_{\mr l}$, we
restrict the summation over the virtual transitions $m \rightarrow
m'$ in Eq.\,(\ref{polar}) to those where both the initial and
final states are elements of the correlated subspace. This is
straightforward in materials where the subspace is formed by
isolated bands so that the partitioning of states is unique.
However, such a case is an exception. In most materials, as in
most of those considered in the present work, the bands forming
the subspace are entangled with other bands, and a clear
separation is not possible. Here, we employ a method outlined in
Ref.\,\onlinecite{cRPA_Sasioglu}. We calculate for each state
$\varphi_{\mathbf{k}m}^\sigma$ the probability
$p_{\mathbf{k}m}^\sigma$ of finding an electron that resides in
that state within the correlated subspace. From
Eq.\,(\ref{Wannier}) it follows that
$p_{\mathbf{k}m}=\sum_{\mathbf{R},n}|T_{\mathbf{R},mn}^\sigma(\mathbf{k})|^2$.
The polarization matrix $P_{\mr l}$ is then calculated from
Eq.\,(\ref{polar}) with the additional factor
$p_{\mathbf{k}m}^\sigma p_{\mathbf{k}^\prime m^{\prime}}^{\sigma}$
for each term of the sum, i.e., for each virtual transition
$\mathbf{k}m \rightarrow {\mathbf{k}^\prime m^{\prime}}$. For
isolated bands, the factor $p_{\mathbf{k}m}^\sigma
p_{\mathbf{k}^\prime m^{\prime}}^{\sigma}$ is simply either 0 or
1, the latter for virtual transitions that take place inside the
correlated subspace, thus comprising the simple case where the
partitioning of states is unique. Yet for the general case of
entangled bands, one has $0<p_{\mathbf{k}m}^\sigma
p_{\mathbf{k}^\prime m^{\prime}}^{\sigma}<1$.

With these definitions, the effective interaction is formally
given by the matrix equation
\begin{equation}
\label{Umatrix}
U = [1-vP_{\mr r}]^{-1}v\,,
\end{equation}
where $v$ is the bare Coulomb matrix. It is related to the fully
screened interaction $\tilde U$, where the screening from the
localized electrons is also taken into account, by
\begin{equation}
\label{Ufullscreen}
\tilde{U} = [1-vP]^{-1}v = [1-UP_{\mr l}]^{-1}U\,.
\end{equation}
The $U$ is nonlocal and inherits a frequency dependence from
$P_\mr{r}(\mathbf{r},\mathbf{r}^\prime;\omega)$. We consider
matrix elements of $U$ in the MLWF basis
\begin{eqnarray}\nonumber
U^{\sigma_1\sigma_2}_{\mathbf{R}n_1 n_3;n_4 n_2}(\omega) & = &
\iint w_{n_1\mathbf{R}}^{\sigma_1*}(\mathbf{r})
w_{n_3\mathbf{R}}^{\sigma_1}(\mathbf{r})
U(\mathbf{r},\mathbf{r}^{\prime};\omega) \\ && \times
w_{n_4\mathbf{R}}^{\sigma_2*}
(\mathbf{r}^{\prime})w_{n_2\mathbf{R}}^{\sigma_2}(\mathbf{r}^{\prime})\:d^3r\:d^3r^{\prime}.
\end{eqnarray}
The average Coulomb matrix elements $U_{\textrm{LDA}+U}$, $U$,
$U'$, and $J$ are defined as follows:
\begin{equation}
U_{\textrm{LDA+}U}=\frac{1}{L^{2}}\sum_{m,n}
U^{{\sigma}_{1}{\sigma}_{2}}_{\mathbf{R}mn;mn}(\omega=0)
\label{Hubbard_U}
\end{equation}
\begin{equation}
U=\frac{1}{L}\sum_{m}
U^{{\sigma}_{1}{\sigma}_{2}}_{\mathbf{R}mm;mm}(\omega=0)
\label{U_diag}
\end{equation}
\begin{equation}
U'=\frac{1}{L(L-1)}\sum_{m \neq n}
U^{{\sigma}_{1}{\sigma}_{2}}_{\mathbf{R}mn;mn}(\omega=0)
\label{U_offdiag}
\end{equation}
\begin{equation}
J=\frac{1}{L(L-1)}\sum_{m \neq n}
U^{{\sigma}_{1}{\sigma}_{2}}_{\mathbf{R}mn;nm}(\omega=0)\,,
\label{Hund_J}
\end{equation}
where $L$ is the number of localized orbitals, i.e., three and
five for \textit{p} and \textit{d} orbitals, respectively. We note
that although the matrix elements of the Coulomb potential $U$ are
formally spin dependent due to the spin dependence of the MLWFs,
we find that this dependence is negligible in practice.
[Henceforth, with $U$ we refer to the average value,
Eq.\,(\ref{U_diag}), rather than to the matrix,
Eq.\,(\ref{Umatrix}).]

Different conventions exist in the literature for the definition
of the Hubbard $U$. (For a detailed discussion see
Ref.\,\onlinecite{Biermann}.) Historically, the Hubbard $U$ was
introduced as a Coulomb repulsion parameter between electrons in
the single-orbital Hubbard-Kanamori-Gutzwiller
model.\cite{Single_band} Note that there is no Hund exchange $J$
for a single orbital. For multi-orbital systems, the Hubbard $U$
is defined as the average intra-orbital and inter-orbital Coulomb
matrix elements. In the present work, we follow the convention
used in Ref.\,\onlinecite{Hubbard_U_def} and  denote the Hubbard
$U$ as $U_{\textrm{LDA+}U}$ and the Hund exchange interaction as
$J$ [see Eqs.\,(\ref{Hubbard_U}) and (\ref{Hund_J})]. As mentioned
before, in contrast to cLDA, the cRPA approach allows to access
individual Coulomb matrix elements and thus, in addition to
$U_{\textrm{LDA+}U}$, we define the average intra-orbital $U$ and
inter-orbital $U'$ Coulomb interaction parameters in
Eqs.\,(\ref{U_diag}) and (\ref{U_offdiag}), which are necessary
for constructing the multi-orbital model Hamiltonians. If the
crystal field has a cubic symmetry, then the $U'$ is given by
$U'=U-2J$. In this case, only two among $U$, $U'$ and $J$ are
independent parameters. In multi-orbital systems, the Hund
exchange $J$ favors spin polarization. Similarly to $U$, $U'$, and
$J$, we can also define the so-called fully screened $\tilde{U}$,
$\tilde{U'}$, and $\tilde{J}$ [see Eq.\,(\ref{Ufullscreen})].
Although the fully screened Coulomb interaction matrix elements
are not used in model Hamiltonians, they provide an idea about the
correlation strength of the considered electrons.

Finally, we would like to note that different conventions of the
Hubbard $U$ parameter in the literature might be confusing for the
reader aiming to use this parameter in LDA+$U$ calculations. In
usual LDA+$U$ methods, the parameter $U_{\textrm{LDA+}U}$ should
be taken as the Slater integral F$^0$. It is also worth to note
that there are two main LDA+$U$ schemes which are in widespread
use today. The Dudarev approach, in which an isotropic screened
on-site Coulomb interaction $U_{eff}=U_{\textmd{DFT}+U}-J$ is
used, and the Lichtenstein approach, in which the
$U_{\textmd{DFT}+U}$ and exchange $J$ parameters are treated
separately.\cite{Dudarev} The Dudarev approach is equivalent to
the Lichtenstein approach for $J=0$.\cite{Dudarev2} Both the
effect of the choice of the LDA+$U$ scheme on the orbital
occupation and subsequent properties like the electronic band
gap\cite{ref5,ref6} as well as the dependence of the magnetic
properties on the value of Hubbard $U$ and Hund exchange $J$ have
been analyzed in the literature.\cite{ref7,ref8,Spaldin} A better
scheme for the LDA+$U$ calculations would be to use the full
Coulomb matrix with a proper treatment of the double counting
issue.

\section{Results and Discussion}\label{sec3}

The results and discussion section is divided into four parts. In
the first part, we present the results of self-consistent
electronic structure calculations to establish the electronic and
magnetic properties of the compounds under study. In the second
part, we present the calculated Coulomb interaction parameters for
all considered systems. The orbital and frequency dependence of
the Coulomb interaction is discussed for selected compounds in the
third and fourth part, respectively. The last part is devoted to
the study of the role that the \emph{p} electrons of the
non-magnetic \emph{sp} element play in the screening of the
Coulomb interactions for three prototype systems.

\subsection{Magnetic moments and  half metallicity} \label{sec3a}

In the present study, we consider a variety of HM magnets, which
are shown in Table\,\ref{table1} together with the structure,
lattice constant, and the calculated spin magnetic moment in units
of $\mu_B$. The first family of compounds are the so-called
\emph{sp}-electron ferromagnets (also known as
\emph{d}$^0$-ferromagnets).\cite{Geshi,Laref} We  consider the
nitrides and the carbides (CaN, SrN, SrC, and BaC) since they have
the largest calculated Curie temperatures among the studied
\emph{sp}-electron ferromagnets.\cite{N1,N2,N3,N4,C1,C2,C3} These
compounds crystallize in the rocksalt structure and do not contain
TM atoms. Electronic structure calculations show that they are
magnetic, and their total spin magnetic moment in units of $\mu_B$
equals $8-Z_{\mathrm{t}}$, where Z$_{\mathrm{t}}$ is the total
number of valence electrons in the unit cell. The latter are
formally made up of the $sp$ states of N or C, while the $s$ state
of the A element (Ca, Sr, or Ba) is located at such a low energy
that it is not classified as a valence state. This rule for the
total spin magnetic moment is known as Slater-Pauling (SP)
behavior and was first identified in the transition-metal binary
compounds.\cite{Slater,Pauling} The number 8 stems from the number
of available $sp$ states: four majority and four minority-spin
states. The former (one of \emph{s} character low in energy and
three bonding \emph{p} states) are all occupied, while the latter
are only partially filled. So, $8-Z_\mathrm{t}$ is the number of
unfilled minority-spin states and equals the total spin magnetic
moment as a consequence. The Fermi level crosses the bonding
minority-spin \emph{p} states, and an energy gap forms in the
majority states between the bonding and antibonding \emph{p}
states. As can be concluded from the spin-magnetic moment
presented in Table\,\ref{table1}, all four compounds under study
are half-metals with a total spin magnetic moment of 1 $\mu_B$ for
the nitrides and 2 $\mu_B$ for the carbides. In the nitrides, the
spin moment is carried mainly by the N atoms, while in the
carbides a large portion of the spin magnetic moment is located in
the interstitial region, i.e., away from the atomic nuclei. As a
reminder, in the FLAPW method the space is divided into
non-overlapping muffin-tin spheres, which are centered around each
atom, and the remaining interstitial region. The muffin-tin radii
of N and C, relevant in this case, are chosen to be about 1~\AA{}
each.

\begin{table}
\caption{Crystal structures (RS stands for rock-salt and ZB for
zincblende), lattice constants, atom-resolved, interstitial, and
total spin magnetic moments (in $\mu_B$) for all considered HM
magnets. Lattice constants are taken from Refs.\,\onlinecite{N3},
\onlinecite{C1}, \onlinecite{landolt},  \onlinecite{APL11}, and
\onlinecite{ZBexch}.}
\begin{ruledtabular}
\begin{tabular}{llcrrrrrr}
 Comp. & Str. & a(\AA) & $m_{\textrm{A}}$ &
$m_{\textrm{B}}$ &  $m_{\textrm{C}}$ &
$m_{\textrm{D}}$ &  $m_{\textrm{Int}}$ & $m_{\textrm{T}}$ \\
\hline
CaN        &RS    &5.02& 0.05 &     & 0.79&     & 0.16& 1.00 \\
SrN        &RS    &5.37& 0.04 &     & 0.81&     & 0.15& 1.00 \\
SrC        &RS    &5.67& 0.13 &     & 1.15&     & 0.72& 2.00 \\
BaC        &RS    &6.00& 0.14 &     & 1.15&     & 0.71& 2.00 \\
VAs        &ZB    &5.69& 1.89 &-0.15&     &     & 0.17& 2.00 \\
CrAs       &ZB    &5.65& 2.99 &-0.25&     &     & 0.26& 3.00 \\
MnAs       &ZB    &5.65& 3.58 &-0.17&     &     & 0.27& 3.68 \\
FeMnSb     &C1$_b$&5.88&-1.14 & 3.13&     &-0.03& 0.04& 2.00 \\
CoMnSb     &C1$_b$&5.87&-0.21 & 3.29&     &-0.09& 0.01& 3.00 \\
NiMnSb     &C1$_b$&5.93& 0.25 & 3.72&     &-0.06& 0.09& 4.00 \\
Mn$_2$VAl  &L2$_1$&5.93&-1.55 & 1.00&-1.55& 0.03& 0.07&-2.00 \\
Mn$_2$VSi  &L2$_1$&5.76&-0.73 & 0.43&-0.73& 0.02& 0.03&-0.98 \\
Cr$_2$FeGe &XA    &5.76&-1.23 & 1.51&-0.27&-0.01& 0.03& 0.03 \\
Cr$_2$CoGa &XA    &5.80&-1.94 & 1.72& 0.42&-0.05&-0.08& 0.07 \\
Co$_2$CrAl &L2$_1$&5.73& 0.80 & 1.55& 0.80&-0.07&-0.08& 3.00 \\
Co$_2$CrSi &L2$_1$&5.65& 1.00 & 2.04& 1.00&-0.05& 0.01& 4.00 \\
Co$_2$MnAl &L2$_1$&5.76& 0.77 & 2.73& 0.77&-0.10&-0.12& 4.05 \\
Co$_2$MnSi &L2$_1$&5.65& 1.05 & 3.01& 1.05&-0.06&-0.05& 5.00 \\
Co$_2$FeAl &L2$_1$&5.73& 1.22 & 2.80& 1.22&-0.07&-0.17& 5.00 \\
Co$_2$FeSi &L2$_1$&5.64& 1.37 & 2.82& 1.37&-0.01&-0.07& 5.48
\end{tabular}
\end{ruledtabular}
\label{table1}
\end{table}

The second family of compounds are the binary VAs, CrAs, and MnAs.
The interest in them started to grow in 2000 when Akinaga and his
collaborators managed to grow a multilayers CrAs/GaAs
structure.\cite{Akinaga} CrAs was found to adopt the zincblende
structure of GaAs. It was predicted to be a half-metal, and the
experimentally determined total spin magnetic moment was found to
be in agreement with this prediction.\cite{Akinaga} Several
studies followed this initial discovery, and electronic structure
calculations have confirmed that also similar binary XY compounds,
where X is an early transition-metal atom and Y an \emph{sp}
element, should be half-metals.\cite{Mavropoulo03,ReviewZB} The
energy gap is located in the minority-spin electronic band
structure and is created from the \emph{p-d} hybridization effect.
The TM \emph{d} orbitals of $t_{2g}$ symmetry transform according
to the same symmetry operations as the \emph{p} valence states of
the \emph{sp} atom, which enables hybridization among them. In the
majority-spin bands, the bonding hybrids are mainly of \emph{d}
character, while in the minority-spin band structure the bonding
hybrids are of \emph{p} character leading (if we also take into
account the single deep-lying \emph{s} valence states) to a
$Z_{\mathrm{t}}-8$ Slater-Pauling rule for the total spin magnetic
moment.

The last family of HM magnets under study are the Heusler
compounds. These systems have been widely studied because the half
metallicity in the bulk samples is
well-established,\cite{FelserRev} and most of them have very high
Curie temperatures approaching or even exceeding 1000
K.\cite{landolt} The first compound that was predicted to be a
half-metal was NiMnSb,\cite{Groot1983} but most of the research
attention during the last years has been focused on the
full-Heusler compounds. This family of materials encompasses many
more members with diverse magnetic properties. There are strong
ferromagnets like Co$_2$MnSi and Co$_2$FeAl, ferrimagnets like
Mn$_2$VAl and Mn$_2$VSi and even HM antiferromagnets like
Cr$_2$FeGe and Cr$_2$CoGa. Especially HM antiferromagnets,
initially predicted by van Leuken and de Groot,\cite{Leuken} are
of interest as they combine half-metallicity with a zero total net
magnetization, which is ideal for spintronics devices due to the
vanishing external stray fields created by them; we should add
that thin films of Cr$_2$CoGa have been grown
experimentally,\cite{Cr2CoGa} a material that has been predicted
to exhibit extremely high Curie temperature.\cite{APL11} The
hybridization that gives rise to the bands responsible for the
formation of the energy gap is complicated and has been
extensively discussed in the literature together with the
resulting Slater-Pauling rules (see Ref.\,\onlinecite{Gala02a} for
half-Heusler compounds, Ref.\,\onlinecite{Gala02b} for the usual
L$2_1$-type Heusler compounds, and Ref.\,\onlinecite{Gala13} for
the inverse XA-type Heuslers). All compounds under study are
half-metals, and the atom-resolved spin magnetic moments are in
good agreement with previously published data. Here we should also
note that Co$_2$FeSi, should have a total spin magnetic moment of
6 $\mu_B$ in case of half-metallicity. Standard GGA calculations
yield a spin magnetic moment of about 5.5 $\mu_B$, and the Fermi
level is below the energy gap. Calculations within the LDA+$U$
method\cite{FelserU} or the $GW$ approximation\cite{BlugelGW}
restore the HM character shifting the Fermi level in the gap. On
the other hand, recent results by Meinert and collaborators show
that a self-consistent calculation fixing the total spin magnetic
moment to 6 $\mu_B$ reproduces more accurately the position of the
band with respect to available experimental
data.\cite{MeinertCo2FeSi}

\subsection{Effective on-site Coulomb interaction parameters}
\label{sec3b}

\subsubsection{Binary compounds}

We start with the discussion of the results of Coulomb interaction
parameters for the \emph{sp}-electron ferromagnets (see
Table\,\ref{table2}). In these compounds, no \emph{d} valence
states are present. The $p$ states of N (C) form bands that are
disentangled from the rest of the band structure with some small
admixture of Ca (Sr, Ba) $s$ states. Hence, these bands lend
themselves for the construction of MLWFs forming the correlated
subspace. In the case of two nitrides, CaN and SrN, the existence
of an energy gap above the Fermi energy (see
Ref.\,\onlinecite{N3}) leads to a less efficient screening of the
\emph{p} electrons and, as a consequence, we obtain larger Coulomb
matrix elements (see Table\,\ref{table2}). The calculated
intra-orbital $U$ exceeds 4 eV for the nitrides, which is
comparable to the values for the elementary TMs presented in
Ref.\,\onlinecite{cRPA_Sasioglu}, while for the carbides the
calculated values are about 1 eV smaller.  The inter-orbital
Coulomb matrix element $U'$ follows the same trend: it is about
3.5 eV for the nitrides and below 3 eV for the carbides. The
behavior of $U$ and $U'$ is reflected also in $U_{\textmd{DFT}+U}$
(Hubbard $U$), which is close to 4 eV for the nitrides and 1 eV
less for the carbides.  We should note here that as we move from
Ca to Sr or from Sr to Ba, although the number of valence
electrons does not change, the increase of the lattice constant
leads to a narrowing of the \emph{p} bands, which gives rise to a
more efficient screening of the Coulomb interaction and
consequently to smaller matrix elements.\cite{Sasioglu_2012_prl}
The value of the Hund exchange parameter $J$ is considerably
smaller than the Coulomb repulsion terms. Its value is around
0.4-0.5 eV for these compounds. As we discussed above, we exclude
the \emph{p}$\rightarrow$\emph{p} transitions in the polarization
function when calculating the partially screened Coulomb
interaction. If we include these transitions, we get the fully
screened Coulomb matrix elements $\tilde{U}$, $\tilde{U'}$, and
$\tilde{J}$, which are much smaller than the corresponding $U$,
$U'$ and $J$ as expected.\cite{cRPA_Sasioglu}

\begin{table*}
\caption{Calculated average partially screened
($U_{\mathrm{LDA+}U}$, $U$, $U'$, and $J$) and fully screened
($\tilde{U}$, $\tilde{U'}$, and $\tilde{J}$) Coulomb interaction
parameters between the localized orbitals denoted in the second
row (in eV) for all HM magnets. We note that unscreened (bare)
Coulomb matrix elements (results not shown) for 3\emph{d} atoms in
HM magnets are similar to the ones in elementary 3\emph{d} TMs
presented in Ref.\,\onlinecite{cRPA_Sasioglu}.}
\begin{ruledtabular}
\begin{tabular}{llccccccc}
Compound & Orbital & $U_{\textmd{LDA}+U}$ & $U$ & $U^{\prime}$ & $J$ &
$\tilde{U}$ & $\tilde{U'}$ & $\tilde{J}$ \\
\hline
CaN        &N-2\emph{p}  &3.90 &4.56 &3.57 &0.52 &0.67 &0.16 &0.27 \\
SrN        &N-2\emph{p}  &3.70 &4.33 &3.38 &0.51 &0.60 &0.13 &0.25 \\
SrC        &C-2\emph{p}  &3.04 &3.52 &2.80 &0.40 &1.01 &0.44 &0.30 \\
BaC        &C-2\emph{p}  &2.67 &3.10 &2.45 &0.37 &0.75 &0.29 &0.26 \\
VAs        &V-3\emph{d}  &1.98 &2.80 &1.77 &0.53 &1.21 &0.43 &0.40 \\
CrAs       &Cr-3\emph{d} &1.67 &2.57 &1.45 &0.57 &1.61 &0.64 &0.48 \\
MnAs       &Mn-3\emph{d} &1.73 &2.63 &1.50 &0.56 &1.87 &0.84 &0.51 \\
FeMnSb     &Fe-3\emph{d} &2.16 &3.10 &1.92 &0.60 &0.88 &0.20 &0.36 \\
           &Mn-3\emph{d} &3.06 &4.02 &2.82 &0.59 &1.56 &0.61 &0.46 \\
CoMnSb     &Co-3\emph{d} &2.31 &3.33 &2.05 &0.64 &1.34 &0.43 &0.46 \\
           &Mn-3\emph{d} &2.80 &3.72 &2.57 &0.58 &1.63 &0.68 &0.47 \\
NiMnSb     &Ni-3\emph{d} &2.70 &3.83 &2.42 &0.70 &2.24 &1.04 &0.60 \\
           &Mn-3\emph{d} &2.65 &3.55 &2.42 &0.56 &2.00 &0.99 &0.50 \\
Mn$_2$VAl  &Mn-3\emph{d} &2.89 &3.84 &2.65 &0.60 &1.13 &0.33 &0.39 \\
           &V-3\emph{d}  &3.06 &3.95 &2.84 &0.55 &1.51 &0.61 &0.43 \\
Mn$_2$VSi  &Mn-3\emph{d} &3.09 &4.10 &2.84 &0.63 &1.19 &0.34 &0.41 \\
           &V-3\emph{d}  &3.32 &4.25 &3.09 &0.58 &1.62 &0.67 &0.45 \\
Cr$_2$FeGe &Cr$^{\textrm{A}}$-3\emph{d}&3.05 &3.98 &2.82 &0.58 &1.18 &0.36 &0.40 \\
           &Cr$^{\textrm{B}}$-3\emph{d}&3.21 &4.21 &2.96 &0.63 &1.03 &0.25 &0.41 \\
           &Fe-3\emph{d} &3.46 &4.56 &3.19 &0.69 &1.16 &0.30 &0.42 \\
Cr$_2$CoGa &Cr$^{\textrm{A}}$-3\emph{d}&3.26 &4.20 &3.02 &0.59 &1.33 &0.45 &0.43 \\
           &Cr$^{\textrm{B}}$-3\emph{d}&3.23 &4.22 &2.98 &0.63 &1.21 &0.36 &0.42 \\
           &Co-3\emph{d} &3.84 &5.02 &3.55 &0.73 &1.41 &0.43 &0.48 \\
Co$_2$CrAl &Co-3\emph{d} &3.74 &4.89 &3.45 &0.72 &1.61 &0.56 &0.50 \\
           &Cr-3\emph{d} &2.82 &3.68 &2.60 &0.54 &0.85 &0.21 &0.32 \\
Co$_2$CrSi &Co-3\emph{d} &3.80 &4.95 &3.51 &0.71 &1.88 &0.76 &0.53 \\
           &Cr-3\emph{d} &2.73 &3.55 &2.53 &0.51 &1.04 &0.33 &0.35 \\
Co$_2$MnAl &Co-3\emph{d} &3.40 &4.53 &3.12 &0.70 &1.64 &0.60 &0.50 \\
           &Mn-3\emph{d} &3.23 &4.17 &3.00 &0.58 &1.58 &0.64 &0.45 \\
Co$_2$MnSi &Co-3\emph{d} &3.28 &4.40 &3.00 &0.70 &1.83 &0.74 &0.53 \\
           &Mn-3\emph{d} &3.07 &3.98 &2.84 &0.57 &1.71 &0.75 &0.46 \\
Co$_2$FeAl &Co-3\emph{d} &3.00 &4.06 &2.73 &0.66 &1.68 &0.66 &0.49 \\
           &Fe-3\emph{d} &3.43 &4.49 &3.17 &0.66 &1.99 &0.89 &0.53 \\
Co$_2$FeSi &Co-3\emph{d} &3.07 &4.11 &2.81 &0.65 &1.70 &0.68 &0.49 \\
           &Fe-3\emph{d} &3.40 &4.43 &3.14 &0.64 &1.33 &0.43 &0.46 \\
\end{tabular}
\label{table2}
\end{ruledtabular}
\end{table*}

In the case of the binary ferromagnets that contain a TM element
(VAs, CrAs, and MnAs), the situation is more complicated. Now, the
correlated subspace is composed of the valence \emph{d} states of
the TM atoms. Due to the tetrahedral symmetry, the \emph{d} states
are split into the triply degenerate $t_{\mathrm{2g}}$ and doubly
degeneraty $e_{\mathrm{g}}$ states. The former transform in the
same way as the valence \emph{p} states of As. So, they can
hybridize and they do so strongly. Therefore, we construct the
five Wannier $d$ orbitals out of eight electronic bands, five $d$
and three $p$ bands. We thus have a case of entangled bands, for
which we employ the procedure outlined in Sec.\,\ref{sec2}. The
$U$ values in VAs, CrAs, and MnAs amount to about 2.6-2.8 eV.
These values are small when compared with the values for the
elementary TMs in Ref.\,\onlinecite{cRPA_Sasioglu}, where we found
them to be about 3 eV for V and around 4 eV for Cr and Mn. The
difference can be attributed to the very efficient screening
produced by the As \emph{p} electrons. Thus, we expect the
correlation to be weak in these compounds and to have only a small
influence on the electronic structure. The Hund exchange parameter
$J$ is slightly above 0.5 eV, while it is around 0.6-0.7 eV in
elementary TMs. Interestingly, taking into account in addition the
screening inside the correlated subspace, i.e., considering
$\tilde{U}$, $\tilde{U}^\prime$, $\tilde{J}$ values, does not
reduce the values as drastically as we have observed in the
\emph{sp}-electron ferromagnets. Finally, we would like to note
that the Hubbard $U$ parameter for CrAs was already calculated in
Ref.\,\onlinecite{ChioncelCrAs} by employing the cLDA method. The
authors obtained a value of 7 eV, which is much larger than our
value of 1.67 eV (see Table\,\ref{table2}). As mentioned in the
introduction, this unreasonably large value obtained within cLDA
can be attributed to the difficulties in compensating for the
self-screening error of the localized electrons (see
Ref.\,\onlinecite{cRPA_2} for a detailed discussion).

\subsubsection{Heusler compounds}

Finally, we will discuss the case of Heusler compounds, which are
the most widely studied HM magnets. Now, there are two different
kinds of 3\emph{d} TM atoms in the unit cell, and the calculation
of the Coulomb matrix elements becomes more heavy. In the
construction of the Wannier functions for the TM atoms, we include
13 bands for the semi-Heusler compounds and 18 bands for the
full-Heusler compounds, taking into account not only the 3\emph{d}
states of the TM atoms but also the valence \emph{p} states of the
non-magnetic \emph{sp} atom. In half-metallic Heusler compounds,
the width of the \emph{d} bands is usually around 6 eV as can be
deduced from several published plots of the density of states
(DOS).\cite{Gala02a,Gala02b} This large value of the band width is
due to the strong hybridization of the \emph{d} and \emph{p}
valence states of the neighboring atoms. It is larger than the
calculated $U$ values presented later in this section, and thus
half-metallic Heuslers should be classified as weakly correlated
materials. We do not present the band structures here since for
most HMs, and Heusler compounds in particular, several
publications have already been dedicated to describe their band
structures in detail.\cite{ReviewHM,Review_DMS,Gala02a,Gala02b}

In the series of the semi-Heusler compounds $X$MnSb with $X$ = Fe,
Co, Ni, where in each substitution the valence \emph{d} electrons
increase by one, we observe a decrease of the $U$ value for the Mn
\emph{d} orbitals from 4.02 to 3.55 eV, while simultaneously the
$U$ value for $d$ orbitals at the $X$ atom increases from about
3.1 eV for Fe to 3.83 eV for Ni (see Table\,\ref{table2}). The
increase of $U$ for the $X$ atom can be qualitatively understood
on the basis of the behavior of the Mn and  $X$ 3\emph{d} states
(see Fig.\,\ref{fig2} for the atom resolved DOS of NiMnSb and
Refs.\,\onlinecite{CoMnSb_dos} and \onlinecite{FeMnSb_dos} for
CoMnSb and FeMnSb). The Fe atom is close to half-filling. In the
DOS of FeMnSb there is thus a strong Fe 3\emph{d} weight around
the Fermi level (see Ref.\,\onlinecite{FeMnSb_dos}) giving rise to
efficient screening of the Coulomb interaction through
\emph{d}$\rightarrow$\emph{sp} transitions; the closer the
occupied and unoccupied states are to the Fermi level, the larger
the polarization (see next section for a detailed discussion). As
the 3\emph{d}-electron number of the $X$ atom increases, the
3\emph{d} weight around the Fermi level decreases and, as a
consequence, the Coulomb interaction parameters increase.  A
similar trend is observed in the matrix elements of the fully
screened Coulomb interaction ($\tilde{U}$, $\tilde{U'}$, and
$\tilde{J}$), which can be explained by the same arguments. In the
case of the elementary 3\emph{d} TM series, the $U$ value shows a
plateau around 4 eV from Cr to Ni, while the presently calculated
values for the $X$ atom are considerably smaller. This indicates
that the crystal field and the hybridization between the orbitals
of the neighboring atoms have a crucial impact on the $U$ values,
considerably altering the screening of the orbitals in question.
The same conclusion stands also for the inter-orbital $U'$
parameter, while the Hund exchange parameter $J$ is much less
affected by the crystal field or the chemical formula of the
compounds.

In the case of the two HM ferrimagnets (Mn$_2$VAl and Mn$_2$VSi),
the $U$ values for the Mn atoms approach that of elementary Mn.
For the non-magnetic elementary bcc V the $U$ value is about 3 eV,
while in the two compounds under study it exceeds 4 eV. This
difference can be easily explained by the spin polarization of the
V 3\emph{d} states in these Heusler compounds. As will be
discussed in the following section and as it has been shown in
Ref.\,\onlinecite{Sasioglu} for the case of Fe, Co, and Ni, the
spin polarization has a strong influence on the screening of the
Coulomb interaction giving rise to larger $U$ values due to the
decreased weight of the 3\emph{d} states around the Fermi level.
The same discussion holds also for these two Heusler compounds.
The Hund exchange $J$ in elementary bcc V is about 0.6 eV, almost
identical to its value in the present two Heusler compounds. The
small difference as we move from Al to Si for all calculated
parameters is due to the smaller lattice constant of the Mn$_2$VSi
compound, which leads to a slightly larger Coulomb repulsion
between the 3\emph{d} electrons.

The rest of the compounds are full Heusler alloys, where the first
two are antiferromagnetic and in the XA structure, while the
others are ferromagnetic and exhibit an L$2_1$ structure. Here,
the calculated $U$ parameters show a more complex behavior. The Cr
atoms exhibit a large variation of $U$ between the XA-type and the
L$2_1$-type compounds, which are about 4.0-4.2 eV in the former
and about 3.5-3.7 eV in the latter. Also, $J$ is reduced from 0.6
eV in the XA-type compounds to 0.5 in the L$2_1$-type compounds.
There are two inequivalent Cr atoms per unit cell in the XA-type
compounds with different chemical environments. Each has eight
nearest neighbors of which four are Cr atoms and the other four
are Fe (Co) atoms in the one case and Ge (Ga) atoms in the other.
In the L$2_1$-type compounds Co$_2$CrAl and Co$_2$CrSi, each Cr
atom has eight Co atoms as nearest neighbors. The different
hybridization with the neighboring atoms influences the values of
the calculated parameters. The Mn atoms in Co$_2$MnAl and
Co$_2$MnSi show $U$ values close to those of the Mn atoms in the
ferrimagnetic compounds discussed in the previous paragraph. The
3\emph{d} orbitals of Fe have a $U$ value of about 4.4-4.5 eV and
a $J$ value of 0.65-0.70 eV nearly irrespective of the chemical
type of the compounds. The larger deviations are observed for the
Co atoms. The $U$ values range from 4.1 eV in Co$_2$FeSi up to 5.0
eV in Cr$_2$CoGa, while $J$ changes about 0.1 eV for Co 3\emph{d}
orbitals between the various compounds. The fully screened Coulomb
interaction parameters do not show as much variation. They are
substantially reduced with respect to the partially screened
values.

So far, we have only focussed on the on-site Coulomb interaction
parameters. Due to metallic screening in TM-based HM magnets, the
calculated nearest-neighbor $U$ values turn out to be negligibly
small, i.e, they lie between 0.1 and 0.4 eV as in the 3\emph{d} TM
series.\cite{cRPA_Sasioglu} On the  other  hand, the situation is
different for \emph{sp}-electron ferromagnets for which we obtain
sizable nearest-neighbor $U$ values (0.5-0.7 eV) for \emph{p}
orbitals. This is because the \emph{p} bands, which form the Fermi 
surface, are isolated in these systems; thus, partial screening is 
not metallic.

Similar to the case of CrAs, previous cLDA calculations of the
effective Coulomb interaction for the 3\emph{d} TMs in NiMnsb and
Co$_2$MnSi resulted in Hubbard $U$ values of about 6
eV,\cite{ChioncelPRB03,ChioncelPRL08} which are found to be too
large to be used in material-specific LDA+DMFT calculations. Thus,
in LDA+DMFT studies addressing the correlation effect in
ferromagnetic NiMnSb and Co$_2$MnSi, the Hubbard $U$ and Hund
exchange $J$ for the 3\emph{d} TMs are chosen to be 3 eV and 0.9
eV,\cite{ChioncelPRB03,ChioncelPRL08} respectively, which is very
close to our calculated values (see Table\,\ref{table2}).

\subsection{Orbital dependence of \emph{U} and \emph{$\tilde{U}$}}
\label{sec3c}

As we have already mentioned above, the values presented in
Table\,\ref{table2} for the TM atoms are the average ones of the
Coulomb matrix elements. The lattice structures presented in
Fig.\,\ref{fig1} exhibit tetrahedral symmetry, and the valence
3\emph{d} states thus separate into the doubly degenerate
e$_{\mathrm{g}}$ and the triply degenerate t$_{\mathrm{2g}}$
states. Within the tetrahedral symmetry group, the former are
lower in energy, while in the octahedral symmetry group it is vice
versa. This is well understood in terms of the relative
orientations of the orbitals in space. For the structures under
study, the e$_{\mathrm{g}}$ orbitals point along the coordinate
axes. Figure\,\ref{fig1} shows that in this direction atoms of the
same kind are relatively far away with a different atom in
between. The nearest neighbors of the same element are in the
direction of the the t$_\mathrm{2g}$ orbitals. As a result, the
intra-orbital Coulomb matrix elements show a variation with the
orbital character. In Table\,\ref{table3}, we present the
calculated $U$ and $\tilde{U}$ values for the e$_{\mathrm{g}}$ and
t$_{\mathrm{2g}}$ orbitals of the TM atoms in NiMnSb and
Co$_2$MnSi. The energetic position of the e$_{\mathrm{g}}$ and
t$_{\mathrm{2g}}$ states are reflected in the calculated Coulomb
matrix elements. The screening for the former is less effective,
and we thus get larger Coulomb matrix elements. We note in passing
that the corresponding bare interaction parameters are very
similar, which rules out that a different spread of the Wannier
functions is responsible for the variation. The difference in the
calculated $U$ values between the two different subsets of the
\emph{d} orbitals is about 0.5 eV for Ni and Co and 0.2-0.3 eV for
Mn. These values are slightly larger than for the elementary TMs
in Ref.\,\onlinecite{cRPA_Sasioglu} where, \textit{e.g.}, for Ni
the difference is only 0.14 eV. However, the difference is still
comparatively small so that we can safely say that the average
values presented in Table\,\ref{table2} will capture the essential
characteristics of the correlations in these compounds. We also
note that if we take into account the full screening, the obtained
$\tilde{U}$ values for the e$_{\mathrm{g}}$ and t$_{\mathrm{2g}}$
orbitals in the heavier Ni and Co atoms are almost identical,
while for Mn a larger difference persist.

\begin{table}
\caption{Partially screened $U$ and fully screened $\tilde{U}$ for
$e_g$ and $t_{2g}$ orbitals (in eV) for the  ferromagnetic NiMnSb
and Co$_2$MnSi compounds. We also show the results for the
non-magnetic state in parentheses.}
\begin{ruledtabular}
\begin{tabular}{lcccc}
 & \multicolumn{2}{c}{NiMnSb} & \multicolumn{2}{c}{Co$_2$MnSi} \\
                   &  Ni         &     Mn      &    Co       &   Mn  \\ \hline
$U$($e_{\mathrm{g}}$)          & 4.08 (4.66) & 3.65 (2.83) & 4.69 (4.47) & 4.17 (3.86) \\
$U$($t_{2\mathrm{g}}$)         & 3.66 (4.22) & 3.48 (2.72) & 4.22 (4.10) & 3.85 (3.62) \\
$\tilde{U}$($e_{\mathrm{g}}$)  & 2.40 (2.25) & 2.10 (0.60) & 1.87 (0.69) & 1.96 (0.34) \\
$\tilde{U}$($t_{2\mathrm{g}}$) & 2.14 (1.81) & 1.92 (0.54) & 1.81 (1.23) & 1.54 (0.76) \\
\end{tabular}
\end{ruledtabular}
\label{table3}
\end{table}

\begin{figure}[t]
\begin{center}
\includegraphics[width=\columnwidth]{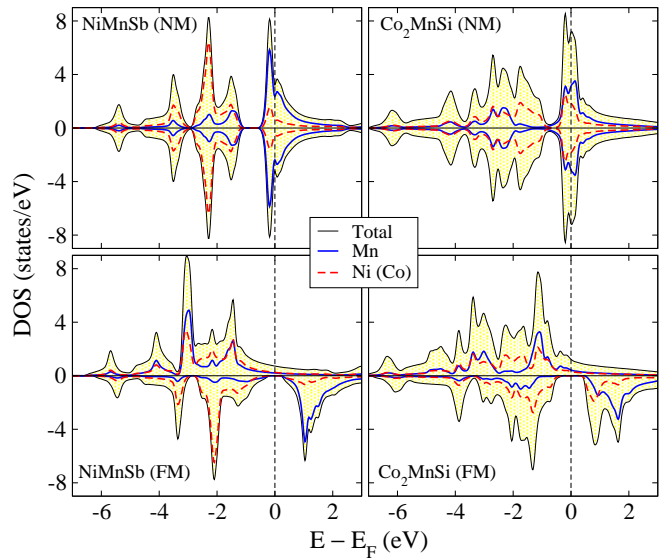}
\end{center}
\vspace*{-0.7 cm} \caption{(Color online) Total and atom-resolved
density of states (DOS) for non-magnetic and magnetic NiMnSb and
Co$_2$MnSi. The zero of the energy axis corresponds to the Fermi
level. Positive (negative) DOS values are associated with the
majority (minority) spin electrons.} \label{fig2}
\end{figure}

In Table\,\ref{table3}, we have also included in parentheses the
values for the case of non-spin-polarized calculations, which
differ substantially from the spin-polarized case, especially the
fully screened Coulomb interaction. A similar behavior has been
observed in the elementary TMs \cite{Sasioglu} and can be
explained with the help of the DOS for the magnetic and
non-magnetic systems presented in Fig.\,\ref{fig2}. First, we
remark that in the non-magnetic calculations the Fermi level is
located in a peak of the DOS. Due to the Stoner criterion, both
NiMnSb and Co$_2$MnSi therefore prefer the ferromagnetic ground
state. The variations in Coulomb matrix elements between the
ferromagnetic and the non-magnetic state (see Table\,\ref{table3})
can be qualitatively explained by the DOS around the Fermi level.
As the screened Coulomb interaction $U$ depends on the
polarizability (see Eq.\,\ref{polar}), the number of occupied and
unoccupied states around the Fermi level plays an important role
in determining its strength. Mn in non-magnetic NiMnSb has the
largest DOS around the Fermi energy and hence the smallest Coulomb
matrix elements with an average $U$ value of about 2.7-2.8 eV. Ni
in non-magnetic NiMnSb, on the contrary, has the smallest DOS
around the Fermi level among the TM atoms in the two compounds and
thus the largest calculated $U$ values. In the non-magnetic state
of Co$_2$MnSi, we observe a density of Co and Mn derived states at
the Fermi level that is in-between that of the Ni and Mn atoms in
NiMnSb (the one of Mn being slightly larger than that of Co),
which is reflected by the calculated $U$ values. For the
ferromagnetic compounds, on the other hand, the corresponding
peaks are shifted to lower and higher energies for the majority
and minority spin, respectively, due to the exchange field,
leading to a lower DOS around the Fermi level. As a consequence,
we obtain larger Coulomb matrix elements in the ferromagnetic
state compared to the non-magnetic state for all TM atoms in
NiMnSb and Co$_2$MnSi.

\begin{figure}[t]
\begin{center}
\includegraphics[width=\columnwidth]{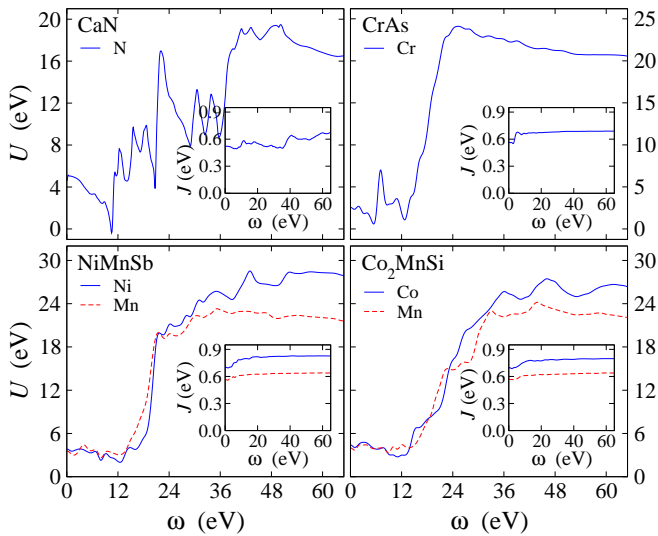}
\end{center}
\vspace*{-0.7 cm} \caption{(Color online) Frequency dependence of
the effective Coulomb interaction parameters $U(\omega)$ and
$J(\omega)$ for the  HM ferromagnetic CaN, CrAs, NiMnSb, and Co$_2$MnSi
compounds.}
\label{fig3}
\end{figure}

\subsection{Frequency dependence of \emph{U} and \emph{J}} \label{sec3d}

The Coulomb interaction parameters presented in
Table\,\ref{table2} are for the static limit, i.e., for $\omega
=0$. The question arises whether the use of the static limit of
the Coulomb interaction in model Hamiltonians is justified. In
Fig.\,\ref{fig3}, we present the frequency dependence $U(\omega)$
and $J(\omega)$ for selected HM magnets: CaN, CrAs, NiMnSb, and
Co$_2$MnSi. In the case of the two Heusler compounds, the
$U(\omega)$ is almost constant at low frequencies up to 15 eV,
suggesting that the use of the static value $U(\omega=0)$ in model
Hamiltonians is appropriate. As we approach the plasma frequency,
slightly above 15 eV, $U(\omega)$ increases rapidly approaching
the unscreened (bare) value, i.e., at high frequencies screening
is not effective anymore. For the binary compounds, the situation
is very different. The frequency-dependent $U(\omega)$ of the Cr
3\emph{d} electrons shows strong variations at low frequencies,
while for the 2\emph{p} electrons of N in CaN the static
approximation fails completely as we have very strong oscillations
of $U(\omega)$ even at low frequencies. In contrast to
$U(\omega)$, the  Hund exchange parameter $J(\omega)$, being
mostly an atomic property, depends only weakly on the frequency
and does not show significant variations at the plasma frequency.
Especially in the case of the \emph{d} electrons, it is almost
constant in the plotted frequency range. This atomic-like behavior
of $J(\omega)$ can be attributed to the fact that the exchange
charge has no $l=0$ component, which makes $J(\omega)$ almost
immune to screening, except at very low frequencies.

\begin{table}
\caption{Effect of the \emph{p} electrons on the screening of the
Coulomb interaction parameters for CrAs, NiMnSb, and Co$_2$MnSi.
In parentheses we show the results from Table\,\ref{table2} for
comparison. The notation  \emph{pd} $\rightarrow$ \emph{pd} means
that \emph{d}$\rightarrow$\emph{d} and
\emph{d}$\rightarrow$\emph{p} transitions are excluded in the
calculation of the polarization function.}
\begin{ruledtabular}
\begin{tabular}{llccc}
Comp.   & Orb. &  $U$(\emph{pd} $\rightarrow$ \emph{pd})  &
$U'$(\emph{pd} $\rightarrow$ \emph{pd}) & $J$(\emph{pd} $\rightarrow$ \emph{pd}) \\
\hline
CrAs       & Cr-3\emph{d}  &  5.67 (2.57) &  4.42 (1.45) & 0.63 (0.57) \\
NiMnSb     & Ni-3\emph{d}  &  5.43 (3.83) &  3.95 (2.42) & 0.74 (0.70) \\
           & Mn-3\emph{d}  &  5.22 (3.55) &  4.04 (2.42) & 0.59 (0.56) \\
Co$_2$MnSi & Co-3\emph{d}  &  5.76 (4.40) &  4.32 (3.00) & 0.72 (0.70) \\
           & Mn-3\emph{d}  &  5.29 (3.98) &  4.12 (2.84) & 0.58 (0.57) \\
\end{tabular}
\end{ruledtabular}
\label{table4}
\end{table}

\subsection{Role of the \emph{sp} atom in the screening} \label{sec3e}

Finally, we discuss the effect played by the non-magnetic
\emph{sp} atom in the screening of the Coulomb interaction between
the TM 3\emph{d} electrons in CrAs, NiMnSb, and Co$_2$MnSi. To
study this effect, we have excluded also the
\emph{d}$\rightarrow$\emph{p} transitions in the calculation of
the polarization function in addition to the
\emph{d}$\rightarrow$\emph{d} transitions. The obtained values for
the Coulomb matrix elements are presented in Table\,\ref{table4}.
For comparison, we have also included the $U$, $U'$, and $J$ in
parentheses from Table\,\ref{table2}. When additionally excluding
the \emph{d}$\rightarrow$\emph{p} transitions, the $U$ values
become similar for all compounds under study and amount to about
5.5 eV. The \emph{d}$\rightarrow$\emph{p} screening channel is
quite efficient in the case of CrAs, where the $U$ increases more
than a factor of two with respect to its value of
Table\,\ref{table2}, from 2.57 eV to 5.67 eV, while $J$ remains
relatively unaffected. In the case of Ni and Mn in NiMnSb, the $U$
values become larger by about 30\% larger when including the
additional screening channel. In Co$_2$MnSi, the difference in $U$
values calculated with and without the
\emph{d}$\rightarrow$\emph{p} screening channel is very small. As
we move from CrAs to NiMnSb and then to Co$_2$MnSi, the number of
3\emph{d} electrons as well as of \emph{s} electrons in the unit
cell increases and, as a consequence, the screening of the
\emph{p} electrons of the non-magnetic \emph{sp} atom is not so
efficient any more. Note that the Coulomb screening is not
additive. An extended discussion of this issue can also be found
in Ref.\,\onlinecite{cRPA_2}.

\section{Conclusions}
\label{sec4}

We have calculated the strength of the effective on-site Coulomb
interaction (Hubbard $U$ and Hund exchange $J$) between localized
electrons in different classes of HM magnets employing the cRPA
method within the FLAPW framework. We have considered (i)
\emph{sp}-electron ferromagnets in rock-salt structure, (ii)
zincblende 3\emph{d} binary ferromagnets, and (iii) ferromagnetic
and ferrimagnetic semi- and full-Heusler compounds. For HM
\emph{sp}-electron ferromagnets, the calculated Hubbard $U$
parameters are between 2.7 eV and 3.9 eV, while for TM-based HM
compounds they lie between 1.7 and 3.8 eV, being smallest for MnAs
(Mn 3\emph{d} orbitals) and largest for Cr$_2$CoGa (Co 3\emph{d}
orbitals). We have found that for the HM full-Heusler compounds
the obtained Hubbard $U$ values are comparable to those in
elementary 3\emph{d} TMs, while for the semi-Heusler compounds the
Hubbard $U$ values are slightly smaller. We have shown that the
increase of the Hubbard $U$ parameter with structural complexity,
i.e., from MnAs to Cr$_2$CoGa, can be attributed to an efficient
screening of the \emph{p} electrons of the non-magnetic \emph{sp}
atoms. The \emph{p} electron screening turns out to be more
efficient for MnAs than for Cr$_2$CoGa.  Our calculated Hubbard
$U$ parameters for CrAs, NiMnSb, and Co$_2$MnSi are about two
times smaller than previous estimates based on the cLDA method.
Furthermore, the band width of the studied compounds are in most
cases smaller than the calculated Hubbard $U$ parameters. The HM
magnets can thus be classified as weakly correlated materials.

The Coulomb interaction parameters play an important role in the
construction of model Hamiltonians aimed to the study of
correlation effects in the electronic structure of HM magnets.
Strong correlations give rise to non-quasiparticle states above
(or below) the Fermi energy at finite temperatures reducing the
spin polarization and, as a consequence, the efficiency of
spintronics devices. We hope that the Hubbard $U$ and Hund
exchange $J$ values presented here will prove helpful for future
LDA+$U$ and LDA+DMFT calculations as well as for other methods
applied to describe correlation effects in HM magnets.

\acknowledgements This work has been supported by the DFG through
the Research Unit FOR-1346.

\end{document}